\documentclass[12pt]{article}
\usepackage{longtable}
\relax
\usepackage{amssymb}

\newcommand{\bo}{{\bar o}}

\def\bo{{\raise.15ex\hbox{\large$\Box$}}}               

\def\face{{\raise.2ex\hbox{$\displaystyle \bigodot$}\mskip-2.2mu \llap {$\ddot
        \smile$}}}                                      

\def\ket#1{\left| #1\right\rangle}              
\def\leftrightarrowfill{$\mathsurround=0pt \mathord\leftarrow \mkern-6mu
        \cleaders\hbox{$\mkern-2mu \mathord- \mkern-2mu$}\hfill
        \mkern-6mu \mathord\rightarrow$}       
\def\dvec#1{\vbox{\ialign{##\crcr
        \leftrightarrowfill\crcr\noalign{\kern-1pt\nointerlineskip}
        $\hfil\displaystyle{#1}\hfil$\crcr}}}           

\def\beq{\begin{equation}}
\def\eeq{\end{equation}}

\def\beqx{\begin{displaymath}}
\def\eeqx{\end{displaymath}}

\def\beql{\begin{eqnarray}}
\def\eeql{\end{eqnarray}}

\newcommand{\bea}{\begin{eqnarray}}
\newcommand{\eea}{\end{eqnarray}}

\def\[{\left [}
\def\]{\right ]}
\def\({\left (}
\def\){\right )}

\def\+{\oplus}

\begin{document}

\hbox{\hskip 12cm NIKHEF/2015-030  \hfil}
\hbox{\hskip 12cm IFF-FM-2015/05  \hfil}
\hbox{\hskip 12cm arXiv:1512.03093  \hfil}

\vskip .3in

\begin{center}
{\large \bf Scan Quantum Mechanics: Quantum Inertia Stops Superposition}

\vspace*{.3in}

{Beatriz Gato-Rivera\footnote{Also known as B. Gato.}} \\

\vskip .2in

{\it Instituto de F\' \i sica Fundamental, IFF-CSIC, \\ Serrano 123, Madrid 28006, Spain \\
Nikhef Theory Group, Science Park 105, 1098 XG Amsterdam, The Netherlands}

\vskip .1in

\end{center}

\begin{center}

\vspace*{0.3in}

{\bf Abstract}

\end{center}

{\small A novel interpretation of the quantum mechanical superposition is put forward. Quantum systems 
scan all possible available states and switch randomly and very rapidly among them. The longer they remain 
in a given state, the larger the probability of the system to be found in that state after a measurement 
is performed. A crucial property that we postulate is quantum inertia, that increases 
whenever a constituent is added, or the system is perturbed with all kinds of interactions.  
Once the quantum inertia $I_q$ reaches 
a critical value $I_{cr}$ for an observable, the switching among the different eigenvalues of that observable 
stops and the corresponding superposition comes to an end, leaving behind a system with a well defined  
value of that observable. Consequently, increasing the mass, temperature, gravitational force, etc. of a 
quantum system increases its quantum inertia until the superposition of states disappears for all the 
observables and the system transmutes into a classical one. Moreover, the process could be 
reversible: decreasing the size, temperature, gravitational force, etc. of a classical system one 
could revert the situation, igniting the switching among the states and leading to the apparent 
superposition again. Entanglement can only occur between quantum 
systems, not between a quantum system and a classical one, because an exact synchronization between 
the switchings of the systems involved must be established in the first place and classical systems 
do not have any switchings to start with. Future experiments might determine the critical inertia $I_{cr}$
corresponding to different observables. In addition, our proposal implies a new radiation mechanism in 
strong gravitational fields, that could contribute to neutron star formation, giving rise to non-thermal 
synchrotron emission. Superconductivity, 
superfluidity, Bose-Einstein condensates, and any other physical phenomena at very low temperatures 
must be reanalyzed in the light of this interpretation, as well as mesoscopic systems in general.}

\vskip .3in

\noindent
25 November 2015 \\ 
(Revised version 16 May 2016)

\newpage

\begin{center}

\vspace*{0.3in}

{\large \bf Contents}

\end{center}

\vskip .5in

1. {\bf Introduction {\ \ } 3}  

\vskip .3in

2. {\bf Scan Quantum Mechanics {\ \ } 4} 

\vskip .15in

{\ \ \ }2.1 Basic ideas {\ \ } 4

{\ \ \ }2.2 Quantum inertia versus quantum switching time {\ \ } 9

{\ \ \ }2.3 Sources of quantum inertia {\ \ } 10

\vskip .3in

3. {\bf Quantum Entanglement {\ \ } 13} 

\vskip .3in

4. {\bf Scan Quantum Mechanics vs other Interpretations {\ \ } 16 }

\vskip .3in

5. {\bf Past, Present and Future Experiments {\ \ } 18 }

\vskip .15in

{\ \ \ }5.1 Interferometry experiments {\ \ } 18

{\ \ \ }5.2 The tourmaline crystal {\ \ } 20

{\ \ \ }5.3 Entanglement experiments {\ \ } 21

{\ \ \ }5.4 Possible experiments to probe Scan Quantum Mechanics {\ \ } 22

\vskip .3in

6. {\bf Conclusions and Final Remarks {\ \ } 24} 

\vskip .3in

{\ \ \ \ }{\bf References {\ \ } 27} 

\newpage



{\it $<<$ If one wishes to provoke a group of normally phlegmatic physicists into a state
of high animation -- indeed, in some cases strong emotion -- there are few tactics better
guaranteed to succeed than to introduce into the conversation the topic of the foundations
of quantum mechanics, and more specifically the quantum measurement problem. $>>$ }

\vskip .3in

Anthony J. Leggett in `The Problems of Physics' \cite{AJL}

\vskip .5in

\section{Introduction}

\vskip .2in

Since its formulation in the 1920s and 1930s, the foundations of quantum mechanics have provided an endless source
of intense and passionate debate and inspiration. Far from obsolete, the interpretation of quantum mechanics 
has become a most vivid and popular topic of scientific inquiry in the last four decades, because the 
conventional Copenhagen interpretation, widely accepted for practical purposes, 
leaves too many questions unanswered. Whereas the microscopic world is 
astonishingly strange, counter intuitive, and very probably non-deterministic by its own nature,
the macroscopic world relies on, to a great extent, intuitive, deterministic laws. 
Measurements of the microscopic entities by classical devices are supposed to produce the 
collapse of the wave functions into one of the possible states, the likelihood of each possible state being 
given by the square of the corresponding coefficient. That is, from all possible states in the superposition 
only one is realized in the measurement process, whereas the others just evaporate, as if macroscopic 
classical devices have the magical power to collapse wave functions. Moreover, the dividing line between 
macroscopic classical and microscopic quantum objects is completely unknown, even though many experiments 
have been performed in the last years trying to approach this frontier. These and some other questions 
have given rise to a plethora of different interpretations of quantum mechanics over the years. 

In this paper we present a novel interpretation of some aspects of quantum mechanics. Our proposal, {\it Scan 
Quantum Mechanics}, is not radically different from conventional quantum mechanics since we let the mathematical 
formalism unmodified, as an effective correct description of the quantum systems. However, we put forward a
mechanism underlying the superposition of quantum states, and a criterion for deciding the validity of 
the quantum description versus the classical description of the physical systems. To be precise,
we postulate a property of all systems, quantum and classical alike, {\it quantum inertia} $I_q$, 
in such a way that quantum behaviour only manifests itself for values of $I_q$ below some critical 
ones $I_{cr}$ (one critical value for each observable). As a result, quantum inertia with its critical 
values marks the dividing line between the quantum and the classical worlds.

In the light of this interpretation, the answers to several long-standing crucial questions and paradoxes 
are most intuitive. This applies especially to the questions related to: the measurement problem,
the superposition of quantum states, the elements of reality, the lack of entanglement between quantum
systems and macroscopic objects, and the dividing line between quantum and classical behaviour, as we
just mentioned. Other less intuitive but unavoidable features 
of this interpretation, like in conventional quantum mechanics, are the lack of well defined physical properties 
for quantum systems, randomness in the output of measurements and nonlocality, not only for quantum correlations 
but also for positions in space (i.e. quantum jumps instead of classical continuous trajectories). 

Nevertheless, Scan Quantum Mechanics can be tested with a number of laboratory experiments -- we propose some 
of them -- and by astrophysical observation, since it implies a new radiation mechanism in strong gravitational 
fields. This mechanism could contribute to neutron star formation and gives rise to non-thermal emission, 
especially $\gamma$-ray synchrotron radiation.

In what follows, as the purpose of this presentation is to emphasize the physical concepts of 
Scan Quantum Mechanics, we will consider only pure states keeping the mathematical formalism to a minimum.
In section 2 we will present the main features of our proposal and we will discuss their applications to shed 
some light on a number of mysterious and controversial aspects of conventional quantum mechanics. Entanglement, 
however, is analyzed separately in section 3. In section 4 we consider Scan Quantum Mechanics versus several other 
interpretations, comparing some of their similarities and differences. 
The most relevant past and present experiments in quantum mechanics are reviewed, and reinterpreted in a much
more intuitive way in the light of our interpretation, in section 5. We also propose some experiments  
in order to determine the critical values $I_{cr}$ of the quantum inertia for some observables. These values
are experimental input, not derivable by current theory, like the masses of the elementary particles in 
the Standard Model of Particle Physics. Finally, section 6 is devoted to conclusions and final remarks.

\vskip .4in

\section{Scan Quantum Mechanics}

\vskip .2in

\subsection{Basic ideas}

\vskip .2in

{\it Scan Quantum Mechanics} consists basically of an interpretation of the wave function in which 
the superposition of the quantum states, at any instant, is only an effective approximate concept. The mathematical 
formalism is the same as in conventional quantum mechanics because there is no need to propose any modifications. For 
simplicity, let us consider a quantum system with only one observable. To be precise, we assume that the system is in
a normalized superposition $ \psi = \sum_{k} c_k \ket{a_k} $ of orthonormal eigenvectors $\ket{a_k}$ of an hermitian
operator $A$, with eigenvalues $a_k$, that represents the observable. Then, the measurement is supposed to collapse 
the initial state, or wave function $\psi$, of the system to a unique well defined state, say $\ket{a_m}$, with the 
particular value $a_m$ of the observable, the probability of encountering this value being given by the Born rule, i.e. 
$|c_m|^2$. Our proposal differs from conventional quantum mechanics in essentially two postulates:

\vskip .3in

1. The quantum systems scan the possible available eigenstates described by the wave function $\psi$ and switch among 
them randomly at very high speed so that, effectively, the systems can be regarded as being in a superposition 
of the different eigenstates $\ket{a_k}$, with eigenvalues $a_k$, at any instant. The relative time the quantum system 
spends in a particular state $\ket{a_m}$ coincides with the probability $|c_m|^2$ of finding that state during
a measurement.

\vskip .3in

2. The quantum switching depends on a crucial property shared by all physical systems, {\it quantum inertia} $I_q$, 
that increases whenever a constituent is added, or the system is perturbed with all kinds of interactions and energies.
If the value of $I_q$ is below the critical one for the observable, $I_q < I_{cr}$, then the quantum switching among 
the available states is at work and the wave function $\psi$ describes the system exactly as in conventional quantum 
mechanics. If, on the contrary, the value of $I_q$ reaches the critical one, $I_q \geq I_{cr}$, then the system is 
unable to switch among the different eigenvalues of the observable, the quantum switching freezes and the 
superposition of states disappears leaving behind a classical system with a well defined value of the observable. 
As a consequence, the system's wave function $\psi$ stops providing the correct description of the system. 

\vskip .3in

The generalization of postulate 2 for quantum systems with more than one observable, $A$, $B$, $C$,... is 
straightforward. Obviously, once the quantum inertia $I_q$ reaches the critical values $I_{cr}^A$, $I_{cr}^B$, 
$I_{cr}^C$,...., for all the observables, the quantum superpositions stop and the system becomes classical, 
with well defined values of all the observables. The situation in between, when $I_q$ has reached one or
more of the critical values of the system, but not all of them, leads to hybrid quantum/classical systems,
similarly as in the case of mesoscopic systems.

Notice that, as long as $I_q < I_{cr}$ for a system, Scan Quantum Mechanics is indistinguishable from conventional
quantum mechanics for all practical purposes although, in principle, it provides a more accurate description of the 
quantum systems at very small time scales for which we lack experimental resolution. Moreover, since the quantum
systems are also evolving unitarily with time, as dictated by the Schr\"odinger equation, the r\^ ole of time 
in Scan Quantum Mechanics is twice as prominent as it is in conventional quantum mechanics. This is no obstacle, 
however, in order to transfer the essential concepts of our proposal to relativistic Quantum Field Theory, as we
will argue later.

For $I_q \geq I_{cr}$,
however, the wave function $\psi$ rather than collapsing goes `out of order', in the sense that it does not provide 
the correct description of the physical systems once the switching among the quantum states has stopped. In other
words, we regard the wave function as only a mathematical description of the system, in agreement with conventional 
quantum mechanics, in the same way that a parabolic trajectory is a description of the motion of a bullet, without 
further physical substance.

A different issue is whether Schr\"odinger's equation should be modified, by adding a non-linear term in the 
Hamiltonian, in order to account for the effect of the quantum inertia on the quantum systems, eventually leading to 
the end of the superposition of states. For us this is not at all obvious because our impression, as we will 
discuss later, is that there are two levels of physical reality: the quantum level and the spacetime level, the 
former being more fundamental than the latter. In our views quantum inertia stays deep inside the quantum level, 
not showing up in the spacetime dynamics explicitly, and its major (or unique) r\^ole is to affect the ability 
of the system to switch among the available quantum states, providing as a consequence a 
criterion for the validity of the wave function and the applicability of the Schr\"odinger equation.

The masses of the particles, the interactions among them and many other perturbations and interactions coming from 
the environment (temperature, gravitation, electric and magnetic fields, collisions,...) all must contribute to 
the load of quantum inertia of a system $I_q$, apart from their contributions to the system's Hamiltonian, either 
to the kinetic or to the potential energy. Consequently, if due to these contributions $I_q$ reaches the critical 
value $I_{cr}$, then the system stabilizes to only one state, say $\ket{a_m}$ with eigenvalue $a_m$. This state will 
be observed as a well defined physical state with value $a_m$ by the measuring devices of the classical world. 

An important feature of the quantum switching versus quantum inertia is its plausible {\it reversibility}. That is, 
if the load of quantum inertia $I_q$ of a quantum system is pushed up until or above the critical value $I_{cr}$
but it can be decreased back below $I_{cr}$, by lowering the 
temperature, the magnetic field, the gravitational force, or by any other means, then as soon as $I_q < I_{cr}$ 
the quantum switching resumes activity leading to the apparent superposition of states again, and the wave 
function $\psi$ comes back to work giving a correct effective description of the system, as a result. 
For the purpose of visualization, one could regard the quantum available states as located at degenerate minima 
of a potential, the quantum system oscillating very rapidly among these minima, like a special type of quantum 
tunneling. The load of quantum inertia $I_q$ raises the barriers among the different states until the switching 
finally stops, when $I_q \geq I_{cr}$, and the system gets trapped in one of the minima as a consequence. The other 
way around, lowering the potential barriers below $I_{cr}$, by decreasing $I_q$, will immediately resume the 
frantic oscillations of the system among the minima of the potential.

Positions in space and momenta are treated as any other observable. To be precise, we propose that the system will be 
switching among available positions in space, without passing through intermediate points, as long as $I_q < I_{cr}$, 
where $I_{cr}$ is the critical quantum inertia that stops the quantum jumps. Therefore, as soon as $I_q \geq I_{cr}$, 
the system will stop the quantum switching among positions and will follow a classical continuous trajectory (momenta 
eigenstates), like a tiny bullet. Quantum jumps are therefore the rule, rather than the exception, in Scan Quantum 
Mechanics, and the possibility that the quantum jumps become continuous classical trajectories, whenever 
$I_q \geq I_{cr}$, opens up a world of new effects 
that have to be taken into account for the study of many physical phenomena. 

We see that in Scan Quantum Mechanics there are more elements of reality than in conventional quantum mechanics, where
there are only `potentialities of being' and, in addition, it is forbidden to formulate such questions. Nevertheless,
these elements of reality last only very short times, as the system is jumping continuously and randomly from one quantum
state to another (and from a position in space to another). Moreover, the quantum states only become well defined physical 
states after all the switchings have stopped and the system has stabilized. 

Due to its relationship with the quantum probabilities, the time spent by the systems in each available state must be 
quantized in terms of a time unit, that we call {\it quantum switching time} $t_S$. Now
one may wonder whether the quantum switching is really random or it follows certain deterministic rules. 
For example, suppose that we have a quantum system in a superposition of three states $\ket{a}$, $\ket{b}$ 
and $\ket{c}$, such that the probabilities to find the corresponding values $a$, $b$ and $c$ by means of a measurement 
are $P(a) = 0.25$, $P(b) = 0.50$ and $P(c) = 0.25$. Does the quantum switching follow a specific pattern, quantified 
in terms of the switching time $t_S$, like $(a,b,b,c,a,b,b,c,a,b,b,c,....)$ or like $(b,a,b,c,b,a,b,c,b,a,b,c....)$? 
If this were the case, then exact knowledge of the state of the system at a given time (if this could be 
possible!) plus exact knowledge of the switching pattern would provide exact knowledge of the resulting physical state 
that one could measure at any subsequent instant, even though the original quantum system lacks a well defined quantum 
state. However, there is no reason a priori for the quantum switching to follow a specific pattern because the only 
requirement is that the times spent in the different states must be proportional to their probabilities as given by 
the corresponding coefficients in $|\psi|^2$. 

As a matter of fact, a specific pattern for the quantum switching would require the existence of hidden variables 
that Scan Quantum Mechanics does not need otherwise and that, in addition, could conflict 
with experimental results, namely with the violation of the Leggett inequalities \cite{Leggett}  
that discard many interpretations of quantum mechanics that make use of nonlocal hidden 
variables\footnote{We remind the reader that the experimental violation of the Bell inequalities discards only
the local hidden variable interpretations of quantum mechanics, i.e. it does not discard the nonlocal ones.}.
Consequently, we conclude that the patterns followed by the quantum switchings must be random, and as a result it is 
not possible to predict which will be the output - the physical state that we can observe - after the switching stops. 
Furthermore, the randomness of the quantum switchings allows us to define a time arrow since randomness is not invariant 
under time reversal. 

Another interesting question, especially relevant when there are many possible states, is whether a given quantum system 
actually switches among all the possible states or only among a subset of them. In our opinion, to switch randomly among 
a subset of all the states is a perfectly consistent possibility for a given quantum system, as the wave function only 
makes complete 
sense, experimentally, when applied to a large ensemble of identical quantum systems. For example, let us consider 
the position of a particle at a time $t$, described by the wave function $\psi(x,t)$ that evolves unitarily according 
to the Schr\"odinger equation. In our proposal, the particle is switching positions extremely fast among a large
number of them, respecting the probabilities given by $|\psi(x,t)|^2$, but not necessarily among all the infinite number 
of positions. The particle, however, does not move back and forth like in a brownian motion with classical trajectories, 
but appears and disappears here and there randomly, without crossing intermediate positions, as was discussed before. 

The foregoing discussion is most helpful in order to understand what is going on when the position of a particle is 
given by a superposition of two trajectories, like in interferometry experiments, either by means of a double 
slit or a beam splitter. The particle then has $50\%$ probability to follow one route or the other. In conventional 
quantum mechanics, due to particle-wave duality the particle follows both routes, showing its wave-like nature, 
until it is detected. Then magically the wave-like nature transmutes into a particle-like nature because in the 
measurement devices nobody sees a wave but just an impact of a particle. In contrast, for Scan Quantum Mechanics the 
particle is a wave-like corpuscle equipped with some wave-like behaviour, which bears little resemblance with classical 
corpuscles and even less resemblance with classical waves, and follows the two routes jumping very fast between them, 
without passing through intermediate positions. 
If there is a double-slit setting, the particle actually passes through the two slits 
unless the thickness of the double-slit screen is extremely thin, as the particle will be able to switch many times 
trajectories while crossing through the slits\footnote{An exception where the particle perhaps crosses through only one 
of the two slits are the screens made of graphene precisely because of their thinness.}. If one or two detectors 
are placed just behind the two slits, in order to determine which slit the particle came across, then obviously the 
particle will impact on only one of the detectors since it will stop switching trajectories (and often even existing) 
afterwards. 

We see that Scan Quantum Mechanics offers a very intuitive explanation to the fact that in interferometry experiments
the particle follows the two trajectories whereas it can be detected in only one. This is the magic of the nonlocality
of the particle's trajectory. This amazing behaviour has also lead some authors to overestimate the observer's r\^ole
in interferometry experiments, as one can notice in statements such as `the observer decides if the particle behaves like 
a particle or like a wave'. We will come back to interferometry experiments in section 5.

To finish the basic ideas of Scan Quantum Mechanics, we will now clarify our views about the   
nature of the wave function $\psi$. Is the wave function something physically real (as proposed 
by de Broglie, Schr\"odinger and Bohm) or only a mathematical description of the system? 
Although Scan Quantum Mechanics is somehow compatible with both views, our personal preference,
as was mentioned before, is that $\psi$ is only a mathematical description of the system. For this reason
we find the expression `collapse of the wave function' unfortunate because nothing is collapsing. Rather, 
the mathematical description of the system given by $\psi$ simply does not work because the system's quantum 
inertia has surpassed the critical value: $I_q \geq I_{cr}$, in the same way that the parabolic trajectory 
of a bullet does not collapse, it simply does not describe the bullet's motion, after it has hit the ground. 
In our opinion, the wave-like behaviour of quantum systems is very probably a direct consequence of the 
fact that elementary particles are excitations of quantum fields, and therefore bear the imprint of a 
perturbation in their very nature. This must be the reason why in `one by one' interferometry experiments the 
quantum systems interfere with themselves (as there are not other systems available to interfere with), 
even though the interference patterns become apparent only for statistical ensembles. As a matter of fact, 
for pedagogical reasons it would be most convenient if quantum systems such as particles, atoms and small 
molecules were not regarded as corpuscles neither as waves but something in between, simply as entities that 
are programmed to behave somehow like waves (for this reason they are described by wave functions) but look more 
like corpuscles, especially when they are catch, although in reality they bear little resemblance with 
classical waves and classical corpuscles.

\subsection{Quantum inertia versus quantum switching time}

\vskip .2in

As was argued before, the time that quantum systems spend in each available state must be quantized in terms of a time unit: 
the quantum switching time $t_S$. To be consistent, $t_S$ must depend on the quantum inertia $I_q$. The simplest guess is:

\begin{equation}
t_S =  C \ \ \frac {h} {I_{cr} - I_{q}} \ , \ \ \ I_{q} < I_{cr} \  , 
\end{equation}

\vskip .1in

\noindent
where $h$ is the Planck constant and $C$ is just a dimensionless proportionality constant. In this expression 
the quantum inertia has dimensions of energy or mass, as we set $c=1$. The classical world corresponds 
to $I_{q} \geq I_{cr}$ where $t_S \rightarrow \infty $. Observe that the minimum value of $t_S$, given 
by $t_{Smin} = C \ h / I_{cr}$ is obtained when the quantum inertia vanishes, unless the latter can reach 
negative values, $I_q < 0$, a possibility that we do not consider. Let us remark that a true 
superposition at any instant would correspond to $t_S = 0$, since then the system would be in all the 
states simultaneously, but this can only occur in the limit of infinite values of $I_{cr}$. 

In the spirit of conventional quantum mechanics, it is tempting to propose as well a lower bound relating the indeterminacy 
on the quantum inertia $\Delta I_q$ with the corresponding indeterminacy on the switching time $\Delta t_S$, such as: 

\begin{equation}
  \Delta I_q \ \Delta t_S \geq h/{4\pi}.  
 \end{equation}  

\vskip .1in

We assume that different observables have different values of the critical quantum inertia $I_{cr}$ and different 
values of the quantum switching time $t_S$, as a result. This we find most natural since, for example, switching 
among positions in space is a very different process than switching among spin states, or photon polarizations. 
If this is the case, quantum and classical behaviour can co-exist in quantum systems and these do not transmute completely 
into classical systems until all the observables have given rise to well defined classical properties. Mesoscopic 
systems are known precisely for displaying mixed quantum/classical effects, and this is explained in Scan 
Quantum Mechanics as produced by the fact that different observables have different values of $I_{cr}$.

In principle, the values of the critical quantum inertia $I_{cr}$ for the observables of the system must be obtained,
or at least estimated, by experiment. This is similar to the case of the masses of the elementary particles, they are
experimental input and so far no theory or model has been able to predict them. Nevertheless, the relation 
$I_{cr} = C \ h / t_{Smin}$ already provides interesting insight into the values of $I_{cr}$ as function of the 
minimum switching time $t_{Smin}$. For example, if we set $C=1$, for $t_{Smin}$ equal to the Planck time 
$t_{Pl} = 5.39 \times 10^{-44}$ s, one gets for the critical quantum inertia 
$I_{cr} = 7.67 \times 10^{19}$  GeV $ = 1.367 \times 10^{-7}$ kg. This value is far too large in order to 
mark the dividing line between the quantum and classical worlds, and therefore should be discarded.

In Table 1 some other values of $I_{cr}$ are given as function of $t_{Smin}$, setting $C=1$. 
We can compare these values with
the masses of molecules currently used in interferometry experiments aiming to find the dividing line with respect 
to the mass and size \cite{Hornberger}. 
For this it is useful to remember that the mass of a nucleon (proton or neutron) is slightly 
below 1 GeV and slightly bigger than an atomic mass unit (amu), used in atomic and molecular 
physics\footnote{The proton mass is $m_p = 0.938272$ GeV while 1 amu $= 0.931494$ GeV.}. 
At present, the largest molecules for which quantum mechanical behaviour has been confirmed are in the mass range
of $10^5$ amu, whereas the smallest objects known to behave according to classical mechanics have a mass of
order $10^{-9}$ kg (or $10^{18}$ amu). As a result, there are 13 orders of magnitude to search for the critical
quantum inertia related to mass. This means 13 orders of magnitude to find the dividing line where the center of 
mass motion of an object obeys quantum mechanics or classical mechanics instead. 

Observe also in Table 1 that the values of $t_{Smin}$ in unambiguous quantum territory, like $10^{-27}$ s corresponding 
to 4407 protons (4439 amu) are orders of magnitude smaller than the smallest lifetime of particle decay. 
This provides a consistency check for our proposal since values of $t_{Smin}$ larger than the lifetime of particles 
could lead to conflict.

\begin{table} 
\centering
\caption{Some values of the critical quantum inertia $I_{cr}$, expressed in GeV and kg, as function of the
minimum switching time $t_{Smin}$ in seconds, using the relation $I_{cr} = h / t_{Smin}$.}
\vskip .5cm
\begin{tabular} {lll} 

\hline

$t_{Smin} ({\rm s})$ & $I_{cr} ({\rm GeV})$ & $I_{cr}({\rm kg})$ \\

\hline

$10^{-42}$ & $4.135 \times 10^{18} $ & $7.37 \times 10^{-9} $ \\
$10^{-40}$ & $4.135 \times 10^{16} $ & $7.37 \times 10^{-11} $ \\
$10^{-38}$ & $4.135 \times 10^{14} $ & $7.37 \times 10^{-13} $ \\
$10^{-36}$ & $4.135 \times 10^{12} $ & $7.37 \times 10^{-15} $ \\
$10^{-34}$ & $4.135 \times 10^{10} $ & $7.37 \times 10^{-17} $ \\
$10^{-32}$ & $4.135 \times 10^{8} $ & $7.37 \times 10^{-19} $ \\
$10^{-31}$ & $4.135 \times 10^{7} $ & $7.37 \times 10^{-20} $ \\
$10^{-29}$ & $4.135 \times 10^{5} $ & $7.37 \times 10^{-22} $ \\
$10^{-27}$ & $4.135 \times 10^{3} $ & $7.37 \times 10^{-24} $ \\

\hline

\end{tabular}
\end{table}

\subsection{Sources of quantum inertia}

\vskip .2in

Now let us discuss in some detail the origin of the different contributions to the quantum inertia $I_q$ of the 
physical systems. These contributions surely affect the Hamiltonian of the quantum systems, and therefore 
the evolution of the wave functions in spacetime as a whole, as solutions to the Schr\"odinger equation. 
The point here is that these contributions also affect the quantum inertia $I_q$, whose domain of 
applicability and influence lies deep inside the Hilbert space, as its value determines the system's 
ability to switch among the different states and has little or nothing to do with the classical 
spacetime at the surface.

We have already considered the masses of the constituents of the physical systems in the previous subsection.
This contribution to the quantum inertia 
$I_q$ is very intuitive and is the major reason behind the dividing line between quantum and classical 
systems, since carefully increasing the number of constituents of a quantum system will eventually lead to its
transmutation into a classical one. The explanation given by Scan Quantum Mechanics to this puzzling 
behaviour is amazingly simple because it reduces to the statement that the quantum inertia load of the system 
has reached or surpassed the critical value: $I_q \geq I_{cr}$. As a consequence, it should be possible to determine 
$I_{cr}$ experimentally by performing interferometry experiments using quantum systems with many atoms, like large 
molecules or tiny crystals, that allow the experimenters to add more atoms gradually, one by one if possible. To
this respect, we recommend the interested reader the excellent review \cite{Hornberger} (dated 2012) of 
nanoparticle interferometry with complex organic molecules and inorganic clusters.

The many perturbations and interactions coming from the environment must
also contribute to the load of quantum inertia of the physical systems. Here we have temperature, gravitational forces, 
electric and magnetic fields and collisions with all kinds of matter and radiation.
One has to take into account that these perturbations affect different observables in different ways 
and also that different observables have, in principle, different values for the critical inertia $I_{cr}$.

As regards temperature, it is well known for decades that in some simple macroscopic systems it is possible to 
induce superpositions of states by lowering the temperature until a few degrees above the absolute zero, for example 
superpositions of magnetic flux with opposite directions. More recently the contrary effect was also shown, for example 
observations of quantum decoherence of fullerens heated at 3000 K in a Talbot-Lau interferometer \cite{Hackermuller}. 
Depending on the interpretation of quantum mechanics at hand, these results are attributed to the decrease or 
increase of thermal noise, allowing the more subtle quantum effects to manifest or to completely dilute, or they 
are explained using environmental decoherence theory or spontaneous collapse models, etc. Although thermal noise 
is ubiquitous, and some decoherence mechanisms might be at work to some extent, Scan Quantum Mechanics offers 
a complementary explanation in order to understand what is really going on. Namely, apart from the thermal noise, 
above certain critical temperature $T_{cr}$ the quantum inertia $I_q$ of the system surpasses the critical value
$I_{cr}$ and, as a result, the system becomes purely classical (with no tunneling possible, for example). Below the 
critical temperature $T < T_{cr}$ the system recovers its quantum behaviour, such as the tunneling ability. 

For very small quantum systems, like atoms, the critical temperature must be very high, so that 
the systems as such wouldn't even exist becoming a lump of plasma instead. But for bigger borderline systems, 
like large molecules and tiny crystals, this mechanism should really be at work, and also in the case of some 
special macroscopic systems. 
For these reasons, we think that all known experimental effects and phenomena that make use of very low 
temperatures must be reanalyzed taking into account the possibility of a critical temperature $T_{cr}$
corresponding to a critical quantum inertia $I_{cr}$. This includes phenomena such as Bose-Einstein 
condensates, superfluidity and superconductivity, which at present are not completely well understood. 
We will come back to this issue in section 5.
 
The gravitational forces have been considered already for a long time as a possible cause for the collapse of the wave
function or for decoherence. In section 4 we will say a few words about these proposals for solving the measurement 
problem. According to Scan Quantum Mechanics, gravity must affect the behaviour of quantum systems in two different 
ways, apart from the corresponding contribution to the potential energy in the Hamiltonian. On the one hand, 
increasing gravitational fields must increase the load of $I_q$ with the corresponding increase in the switching 
time $t_S$, like any other interactions will do. It is therefore possible that the quantum switching of a system 
stops altogether provided the gravitational field reaches a strong enough value. This has very interesting 
implications for astrophysical objects with strong gravitational fields, providing an additional mechanism 
that contributes to neutron star formation, complementing the known ones. Namely, the atoms or molecules immersed 
in such strong gravitational fields could become sort of `classical' in the sense that the electrons could 
stop switching among the positions in the electronic cloud, following classical continuous trajectories instead. 
This would lead to the collapse of atoms and molecules due to the emission of $\gamma$-ray synchrotron radiation 
and provides an efficient mechanism for the capture of electrons by the protons in the nuclei, turning the 
atoms and molecules into neutrons as a result. 


On the other hand, strong gravitational fields may slow down dramatically the evolution of all physical systems with 
time, as perceived by external observers. This must include the quantum switching as well.
As a result, it could be possible, at least in principle, to observe from a 
distance the quantum switching among different states provided the dilation of $t_S$ were important enough. 

Collisions with matter or radiation can also stop the switching of the quantum systems and even destroy them altogether. 
This is the case of particle detectors and most devices that measure quantum systems, called generically observers. 
Collisional decoherence has been studied quantitatively using a Talbot-Lau interferometer \cite{Horn2003} by gradual
admision of different gases into the vacuum chamber and it was found that at room temperature a single scattering event
per molecule of fullerens suffices to fully destroy the interference. As before, we think that these experiments
must be reanalyzed taking into account the possibility of a critical quantum inertia $I_{cr}$.

Some interpretations of quantum mechanics give a crucial r\^ole to conscious observers (humans and animals), the more 
extreme ones claiming that the human consciousness is necessary for the existence of the whole universe as we know 
it\footnote{Fortunately, these extreme anthropocentric views of quantum mechanics are out of fashion nowadays.}. 
According to Scan Quantum Mechanics, the interaction of a quantum system with an observer (with or without consciousness) 
is only a particular type of interaction that may increase the system's load of $I_q$ above the critical value. 
If this is not the case, and under the action of the observer the quantum inertia still stays below the critical value, 
$I_q < I_{cr}$, then the observer has failed to turn the superposition of states into a single well defined state, 
described as `collapse' of the wave function in those interpretations. 

An important remark here is that in some experimental settings the researchers involved do not plan to destroy the 
system, usually a particle, in the measurement process, but only to get some information about it, like the trace it 
leaves in a bubble chamber, for example. Some of these procedures are called {\it weak measurements} and the 
experiments usually involve a weak measurement for the momentum of the particle followed by a normal ({\it strong}) 
measurement for the position of the particle, crashing it against a screen or a detector. In our opinion, the weak 
measurements could disrupt the quantum systems much more than the experimenters are aware of, for example stopping the 
quantum jumps among the available positions, so that the system continues its way as a tiny bullet, following a 
continuous classical trajectory that can be confused with a bohmian trajectory (since this is also a continuous 
classical trajectory). We will say a few more words about these experiments in section 5.

\vskip .4in

\section{Quantum Entanglement}

\vskip .2in

In the light of Scan Quantum Mechanics, it is clear that for two, or more, quantum systems 
to be entangled they must switch among the available states in a synchronized manner according to 
the specific type of entangled property. The typical example is the zero spin particle that decays into two particles, 
both in a superposition with $50\%$ probability of spin up and $50\%$ probability of spin down with respect to a given axis. 
These particles are entangled in such a manner that the measurements of their spins will always give opposite values because 
of conservation of angular momentum. In addition, in our interpretation both particles are switching randomly between the 
two possible states, spin up and spin down, and necessarily synchronized. For example, suppose the switching of the spin of 
one of the particles could start with the pattern $(++-+--+-+-+---++-++-.....)$, in units of the quantum switching time $t_S$, 
then the pattern of the other particle's spin should start as $(--+-++-+-+-+++--+--+.....)$. If a perturbation stops this 
switching between the spin states in one of the particles, for example because it is detected giving the $up$ measurement, 
then the switching of the other particle also stops at exactly the same instant without the need of any signals. This is due to
the fact that the resulting quantum system acts as a whole, independently of the physical locations of its parts in spacetime. 
In other words, the gain of quantum inertia $I_q$ by any of the subsystems adds to the total load of $I_q$ for the complete
system. In the case at hand, the
second particle could in principle continue its existence as a quantum system with one of its properties frozen; that is, 
with well defined spin component $down$ with respect to the given axis, until the release of quantum inertia $I_q$ allows
the particle to switch again between the two spin states, liberating itself from the {\it entanglement bound}. This occurrence 
would signalize the end of the entanglement, as a consequence.

We see again that quantum systems defy our most rooted classical intuition producing very sharp nonlocal effects in the 
classical world. Over the years there have been intense and continuous debates about the nonlocality of quantum 
correlations, being the EPR article and the Bell's inequalities the most famous contributions. There is a recent wave-front 
from those who resist nonlocality in quantum entanglement, this time by invoking quantum wormholes that allow faster-than-light 
signals connecting the entangled particles. 

For Scan Quantum Mechanics, however, there is no "spooky action at a distance", as Einstein liked to refer to this nonlocality, 
because the switchings of the entangled particles are exactly synchronized and, consequently, all of them stop simultaneously at 
the same instant with respect to the quantum time in the Hilbert space. A different issue is how different observers in various 
inertial systems in spacetime would perceive the simultaneous ending 
of the quantum switchings of the particles (if these observations could be possible at all, of course!). According to 
Special Relativity, different observers would see the switchings ending at different times, depending on their relative 
velocities with respect to the particles, some could even see the switching of the surviving particles stopping before
the less fortunate particle crashes.

Therefore it is meaningless to try to identify a physical action in spacetime from one of the 
particles to the others in order to correlate their entangled properties, because although the origin of the situation was the 
crashing of one of the particles, the reaction on the switchings of the particles involved comes from the quantum system 
as a whole, not by parts, as we said before. Thus the conceptual problem with the spooky action at a distance is that
there is no distance, truly, in the quantum system of the entangled particles, nonlocality is only an illusion created by 
spacetime. It seems that the Hilbert space of the quantum states is more fundamental than spacetime\footnote{I thank the 
late Yuval Ne'eman for giving me this suggestion during the Strings 1987 conference in Maryland, USA.}. Therefore 
there is no problem with nonlocality in spacetime as long as there is `locality' in the Hilbert space; i.e. that the
quantum states producing apparent nonlocality in spacetime are however located together in the same wave function.

Another cumbersome aspect of quantum entanglement is its supposed proliferation and ubiquity, contrary to our daily experience 
of the world around. This viewpoint dates back to Schr\"odinger and Von Neumann and, in our opinion, 
relies on a probably true fact and a wrong assumption. The probably true fact is that the evolution of the quantum systems is 
linear, as dictated by Schr\"odinger's equation, and the wrong assumption is that macroscopic classical systems should also be 
treated quantum-mechanically\footnote{Bohr and Heisenberg, 
among others, had the contrary opinion: that macroscopic classical systems were not quantum mechanical 
objects, and therefore they should not be treated as such. The problem is that they did not present any convincing arguments 
of where were the limits between the quantum and the classical worlds. This was in fact the origin of the Schr\"odinger's cat 
gedanken experiment, proposed in order to ridicule the superposition of states in the Copenhagen interpretation, contrary 
to Schr\"odinger's beliefs in a physically existing wave function. The viewpoint that macroscopic classical systems should also be 
treated quantum-mechanically is, however, prevalent nowadays, and is assumed in most interpretations of quantum mechanics.}, 
i.e. they also have a wave function, no matter how complicated, that evolves unitarily with Schr\"odinger's equation. 
With these two ingredients it is straightforward to see that, starting with a quantum system in a superposition, 
any microscopic or macroscopic system that happens to interact with it becomes entangled with it, and this entanglement
spreads irremediably as more and more interactions take place with the environment, giving rise to very long `Von Neumann
chains'. However, from the discussion in the previous paragraph, it is obvious that in Scan Quantum Mechanics 
classical systems cannot get entangled with quantum systems by interacting with them since classical systems do not 
posses any quantum switching to start with, and without the synchronization of the switchings of the systems involved
entanglement is simply not possible. 

Another way to realize the impossibility of entangling quantum systems with classical systems is via quantum inertia, that for 
classical systems such as macroscopic daily life devices, bottles, cats, etc.... is much higher than the critical value. As a 
consequence, trying to entangle a quantum system with one of these macroscopic classical systems is like trying to lift an elephant 
with the help of a butterfly pulling the elephant's back. An exception to this rule would be the situation where the value of
the quantum inertia $I_q$ of the classical system is close enough to its critical value $I_{cr}$, so that the interaction
with the quantum system `manages' to release some amount, decreasing $I_q$ below $I_{cr}$ as a result. 
Only in these cases the entanglement of a quantum system with a classical one would be allowed. 

We conclude that in Scan Quantum Mechanics the Von Neumann chains are very short, as generically they only involve quantum systems that 
were already in superpositions before the interactions took place. For example, the minimal version of the Schr\"odinger's cat chain:
quantum system $-->$  poison bottle $-->$  cat, 
reduces to only the quantum system, and neither the poison bottle nor the cat are in superpositions of any kind, as they 
are macroscopic classical objects. In other words, neither the bottle is in the superposition of $\ket{broken}$ and 
$\ket{unbroken}$ states, nor the cat is in the superposition of $\ket{alive}$ and $\ket{dead}$ states, in agreement with the 
Copenhagen conventional interpretation of quantum mechanics.

The question now arises whether it is possible to distinguish Scan Quantum Mechanics from conventional quantum mechanics in
experimental settings of entangled systems. This is the case of experiments aiming to probe Bell's inequalities 
or Leggett's inequalities, in order to distinguish conventional quantum mechanics from local hidden variable theories 
or nonlocal hidden variable theories, respectively. The answer to this question is that, in principle, it is possible to 
distinguish our interpretation from conventional quantum mechanics in such settings, as it predicts the end of the 
entanglement bounds due to the release of quantum inertia to the environment (to empty space at least). 
For all practical purposes, however, the answer could be negative if very long distances (interstellar or larger)
were necessary in order to test this effect. 

Nevertheless, our interpretation provides a more intuitive picture of what is going on in entanglement than the explanation 
offered by conventional quantum mechanics, especially in the situations where the `surviving' particles are measured 
a long time after the entangled partner was detected. To see this, let us compare the different descriptions 
of entanglement provided by Scan Quantum Mechanics and by conventional quantum mechanics 
in the example of the two entangled particles in the superposition of spin up and spin down states with respect
to a given axis. In conventional quantum mechanics, after the first particle is detected with, say, the $up$ state, 
the other particle entangled with it can remain in the superposition with $50\%$ probability of spin up and $50\%$ probability 
of spin down essentially forever, i.e. with no limitation in time, until it is detected. Then, magically, 
when it is detected, the particle decides to behave appropriately showing the spin $down$ to the measuring apparatus. 
Scan Quantum Mechanics is less magical, since once the first particle is measured with the spin $up$, the synchronized 
switching of the spin in the entangled particle freezes, remaining $down$ for some time, exactly until enough quantum
inertia $I_q$ is released allowing the particle to switch again between the two spin states. 

To finish, let us notice that some entanglement phenomena like sudden death, sudden revival and sudden birth might 
bear some relation with the quantum inertia of the systems.

\vskip .4in

\section{Scan Quantum Mechanics vs other Interpretations}

\vskip .2in

In the next paragraphs we will say a few words about the similarities, differences and possible compatibilities of Scan Quantum 
Mechanics with some other interpretations of non-relativistic quantum mechanics. We recommend the interested reader the short 
panoramic views given in refs. \cite{IQM} and \cite{Bassi} about different interpretations of quantum mechanics, 
the latter about models of wave function collapse. For a humoristic sketch see \cite{Cat}. 
 
{\it Environmental decoherence} due to entanglement overdose is only partially compatible with Scan Quantum Mechanics because 
it assumes that macroscopic classical systems behave quantum-mechanically, i.e. they are described by wave functions,
and their interactions with the environment make practically impossible to identify interference phenomena \cite{Joos}, 
\cite{Zurek} (see also \cite{Hornberger}). 
Very recently, gravitation has been proposed to solve the measurement 
problem through decoherence due to time dilation \cite{Pikovski}, but this proposal is currently receiving much 
criticism  \cite{Sudarsky}, \cite{Diositd}, \cite{Zeh}.
In our interpretation, macroscopic classical systems 
do not behave quantum-mechanically due to their quantum inertia load surpassing the critical value: $I_q \geq I_{cr}$.
The mechanism of environmental decoherence is however compatible with Scan Quantum Mechanics when applied to truly
quantum systems with $I_q \leq I_{cr}$. In this case, however, both effects -- the increase of $I_q$ due to the 
interactions with the environment and the decoherence mechanism -- will contribute for the final loss of coherence.
These mixed effects could well be behind the phenomenon of entanglement sudden death \cite{EntDeath}.
 
{\it Hidden variables} \cite{Bohmhv} could be compatible with Scan Quantum Mechanics if, 
and only if, the quantum switchings 
among the states followed exact sequences. But one has to conjecture a convincing reason for this behaviour, and we
do not find any. In addition we believe in the inherent randomness of the quantum world and therefore we do not 
adhere to the possibility of hidden variables in our proposal. Moreover, probably a hidden variable  
version would be ruled out observationally due to the Leggett inequalities \cite{Leggett} that discard most nonlocal 
hidden variable interpretations of quantum mechanics. 

{\it The collapse of the wave function due to consciousness}, or mental abilities, could only be compatible with Scan 
Quantum Mechanics if these would be capable of increasing the quantum inertia of the systems until or
above the critical value $I_{cr}$, like any other interaction. Since the current understanding of consciousness
 -- the awareness of one's own existence -- and mental abilities is rather scarce, we cannot discuss 
this issue any further. Nevertheless, let us remind that although many aspects of the mind can be built in computers 
and other machines leading to artificial intelligence, the emergence of consciousness just from complex matter 
and interactions is very amazing. This has lead some scientists to conjecture that consciousness could be an 
elementary property of matter, like mass or electric charge, and has lead some others to propose that consciousness 
is non-physical at all \cite{PenShM}, \cite{PenENM}, \cite{Bohm}, \cite{Wigner}.  

{\it The quantum histories approach} \cite{KentH} agrees with Scan Quantum Mechanics in that measurements have to be 
treated as any other interaction. This is all the resemblance, however, as this interpretation also treats classical 
macroscopic systems as quantum objects and has many other sharp differences. 

{\it Collapse Models} \cite{Bassi}.
Now let us consider the proposal to replace Schr\"odinger's equation with a nonlinear version of it, which gives rise to 
several interpretations of quantum mechanics, known generically as collapse models. 
As was explained in section 2, Scan Quantum Mechanics needs not invoke a nonlinear Schr\"odinger's 
equation, but this does not mean that nonlinear versions of Schr\"odinger's equation are fully incompatible with it. 
As we said, one may wonder whether the effect of the quantum inertia $I_q$ should show up manifestly in spacetime,
giving rise to a nonlinear Schr\"odinger's equation. But our impression is that quantum inertia only affects the 
quantum switching among the states in the 
Hilbert space, its action being confined there, not interfering with the dynamics in spacetime.
We must point out, however, that some interpretations of quantum mechanics making use of nonlinear Schr\"odinger's equations   
are truly incompatible with the essence of Scan Quantum Mechanics. For example, the proposals that incorporate spontaneous 
sudden collapse of the wave function $\psi$ in this manner \cite{GhRW} \cite{Bassi}.
In our interpretation, the quantum-to-classical transition is 
far from spontaneous and due solely to the system's quantum inertia surpassing the critical value: $I_q \geq I_{cr}$. 
Moreover, in the event that an amount of $I_q$ is released afterwards, pushing $I_q$ down below the critical value, 
$I_q \leq I_{cr}$, then $\psi$ is expected to describe the system again as the quantum switching among 
the superposition states resume activity. This reversibility is an essential property of Scan Quantum Mechanics, 
unless the system is destroyed by the perturbations or in the measurement process, obviously.  

{\it Gravitation}. 
Another particular class of interpretations that make use of nonlinear versions of Schr\"odinger's equation are those that 
signalize gravitation as the major (or only) cause for the collapse of the wave function \cite{PenrosG}, \cite{Penrose},
\cite{Gao} (see ref. \cite{Singh} for a brief survey). For example, in Penrose's proposal, the collapse of $\psi$ is 
always a gravitational phenomenon. These interpretations take into account the system's own gravitational field in 
the Hamiltonian, resulting in the nonlinear Schr\"odinger--Newton equation \cite{Diosi}, where the gravitational 
self-interaction term depends on $|\psi|^2$ and leads to the collapse of $\psi$. 
For Scan Quantum Mechanics gravity is like any other interaction, 
be it from external sources of gravitation or for the system's own self-interaction. Thus if its action 
on a quantum system does not increase the load of quantum inertia $I_q$ above the critical value $I_{cr}$, then 
gravitation fails to stop the superposition of states, i.e. to `collapse' (to disable rather) the system's wave 
function $\psi$. As a matter of fact, we believe that only very strong gravitational fields, like those produced by neutron 
stars and black holes, are able to accomplish this task, although less strong ones together with some other interactions 
(thermal bath, etc.) could also produce the same result.

\vskip .4in

\section{Past, Present and Future Experiments}

\vskip .2in

Now we will discuss in the light of Scan Quantum Mechanics some important experiments that have been made 
over the years in order to test the conventional quantum mechanics. 

\subsection{Interferometry experiments}

\vskip .2in

Let us start with interferometry experiments. The double-slit experiment 
has been running for more than two centuries by now, designed in 1801 by T. Young in order to investigate the 
wave nature of light. For a very long time, the experiment and its variations made use of photons exclusively. 
In 1961, for the first time, electrons were used as projectiles, followed some years later by neutrons, then
by atoms, and finally by molecules. 
The tendency to apply the experiment to bigger, `more macroscopic' quantum systems has continued until 
the present, with the use of increasing larger molecules involving thousand of atoms.

As we pointed out in section 2, our proposal can help us visualize the behaviour of a particle moving in superpositions 
of two trajectories, either by means of a double-slit setting or a beam splitter. First of all, remember that the 
propagation of the particles along each trajectory has wave-like properties, although this becomes only evident if large 
numbers of particles are involved. In other words, single particles do not create interference patterns, neither diffraction 
nor other wave-like patterns, but just single marks at their arrival at the appropriate screens or detectors. 
Nevertheless, these 
marks seem to have some knowledge about the wave-like underlying nature of the particle, in the sense that the particle  
`interferes' with itself, since it cannot do it with any other subsequent particles, in order to create the interference 
patterns made by many collective impacts and observed at the meeting points of the two trajectories.

Scan Quantum Mechanics states that the particle switches very fast between the two wave-like trajectories, 
without passing by intermediate positions. Therefore, if there is a double-slit screen, the particle actually passes through 
the two slits, unless the thickness of the screen is extremely small, because the particle will be able to switch many times 
trajectories while crossing through the slits. If two detectors are placed behind the two slits, the particle will crash on 
only one of them, obviously, whereas if only one detector is placed behind one of the slits the particle may or may not
crash against it. If the particle manages to avoid the detector blocking one of the trajectories, then it will finally 
crash against the second screen coming from the other trajectory, with no memory about the initial superposition of the two 
trajectories. Finally, if there are no detectors in between the double-slit screen and the second screen, then the particle 
will impact on the latter leaving an impact behind, the accumulation of many such impacts producing an interference pattern. 

As a general rule, in interferometry experiments the superpositions can also involve some other properties of the particles 
besides the trajectories. For example, in the case of neutrons, interferences are observed in perfect agreement with conventional 
quantum mechanics when the superpositions are made between different spin orientations too, and even between different gravitational 
pulls if there is a gravitational gradient between the two trajectories, as was already observed in 1975 \cite{Colella}.  
Scan Quantum Mechanics tells us in these cases that the neutrons are switching between the two trajectories, on the one hand, 
and between the two spin orientations or the two gravitational pulls they feel, on the other. 

In 2011, using {\it weak measurements} \cite{Weakmeas} on the momenta of the particles followed by normal, strong measurements 
of the positions of the particles, the Toronto group of Steinberg claimed to have reproduced the average trajectories 
of single photons in a double-slit interferometer. For this experiment \cite{Steinberg} they used about 30.000 
single photons impacting on a screen placed gradually farther behind the two-slit screen. The trajectories they obtained 
seem to coincide with the continuous classical trajectories of the Bohm interpretation of quantum mechanics \cite{LibrosMiret}
and they apparently answer the question from which slit the photon passed through. However, the 
usefulness and reliability of weak measurements are subject of much controversy\footnote{Unlike the results of strong
measurements, weak values are not constrained to lie within the eigenvalue spectrum of the observable being measured. In
addition, weak-value probabilities can take negative values and measured weak values may be anomalously large, like
100 for the spin of an electron and about 100 for the effective photon number in one arm of an interferometer even
if the entire interferometer contains only one photon.},
and the authors themselves gave a warning: `For the experimentally reconstructed trajectories for our double slit, 
it is worth stressing that photons are not constrained to follow these precise trajectories; the exact trajectory of
an individual quantum particle is not a well-defined concept'. Scan Quantum Mechanics can also shed some light on
the fact that the trajectories found in this experiment seem to coincide with classical continuous trajectories. Namely,
the weak measurements of the particles, supposedly to produce only slight modifications on the momenta, could not be
so weak after all and stop the switching of the particle between the two trajectories, due to the increase of its quantum 
inertia above the critical value, forcing the particle to follow a classical continuous trajectory.

Interferometry experiments using beam splitters are in many cases equivalent to the ones using two-slit screens. One
only needs a beam splitter, two mirrors in order to make the split trajectories to converge again, and appropriate 
detectors. Eventually, other devices are placed intercepting the trajectories, depending on the experiments at hand. For 
example, a weak measurement can be accomplished with a thin piece of birefringent calcite that changes the polarization 
of the photons passing through. As before, the increase of quantum inertia of the particles due to these 
weak measurements could eventually stop the switching of the particles between the two routes. Then the particles will 
continue following a continuous classical trajectory until they can release the excess of quantum inertia, resuming the 
switching between the two trajectories, or until they impact against the detector.

\subsection{The tourmaline crystal}

\vskip .2in

The behaviour of polarized light passing through a tourmaline crystal provides another beautiful example where Scan 
Quantum Mechanics improves our intuition and understanding about what is actually going on inside the crystal.  
Tourmaline crystals have the property of letting through only light plane-polarized perpendicular 
to its optic axis. As a result, if the incident light is polarized perpendicular to the optic axis, it will all go through; 
if parallel to the axis, none of it will go through; while if polarized at an angle $\alpha$ to the axis, a fraction 
$sin^2 \alpha$ will go through. 

Dirac discusses this behaviour of the light passing through tourmaline crystals
in his book `The principles of Quantum Mechanics' \cite{Dirac} and wonders how one can 
understand these results in terms of polarized photons: `This picture leads to no difficulty in the cases when our 
incident beam is polarized perpendicular or parallel to the optic axis. We merely have to suppose that each photon 
polarized perpendicular to the axis passes unhindered and unchanged through the crystal, while each photon polarized 
parallel to the axis is stopped and absorbed. A difficulty arises, however, in the case of the obliquely polarized
incident beam. Each of the incident photons is then obliquely polarized and it is not clear what will happen to such
a photon when it reaches the tourmaline'. Then Dirac explains that the obvious experiment is to use single photons
and to observe what appears on the back side of the crystal: `According to quantum mechanics the result of this experiment
will be that sometimes one will find a whole photon, of energy equal to the energy of the incident photon, on the back
side and other times one will find nothing. When one finds a whole photon, it will be polarized perpendicular to the optic 
axis. One will never find only a part of a photon on the back side. 
If one repeats the experiment a large number of times, one will find the photon on the back side in a fraction 
$sin^2 \alpha$ of the total number of times. Thus we may say that the photon has a probability $sin^2 \alpha$ of passing
through the tourmaline and appearing on the back side polarized perpendicular to the axis and a probability $cos^2 \alpha$
of being absorbed'. 

The description offered by Scan Quantum Mechanics to the passing of light through a tourmaline crystal differs substantially 
from the picture held by conventional quantum mechanics, given by Dirac. First of all, when a photon passes through a crystal
its quantum inertia will necessarily increase, and the exact increment could depend on a number of factors such as the 
crystal structure/pattern, crystal chemical composition, its temperature, the incident angle of the photon and its polarization. 
Thus the picture that the photons polarized perpendicular to the axis pass unhindered and unchanged through the tourmaline crystal 
is not exactly correct, even though these photons are recovered out of the crystal as if nothing had happened to them.   
Second, if the crystal is birefringent, like tourmaline or calcite, 
the polarization of the incident photon may split into a superposition of two perpendicular polarization states with different 
probabilities. Scan Quantum Mechanics then states that the photon will switch very fast and randomly between the two 
polarization states, staying longer in the state with larger probability, and increasing its quantum inertia $I_q$ 
in the process, until it eventually reaches a critical value $I_{cr}$. If this is the case, and the value $I_{cr}$ is
reached and surpassed inside the crystal, then the switching between polarization states stops and the photon continues 
its way in a well defined polarization state. In the case of tourmaline crystals, if the switching stops when the photon
is polarized parallel to the optic axis, then the material absorbs the photon and nothing is observed on the back side of the
crystal, whereas if the switching stops when the photon is polarized perpendicular to the optic axis, then the photon reaches 
the back side of the crystal.
 
\subsection{Entanglement experiments}

\vskip .2in

Now let us consider briefly some experiments performed in order to confirm the quantum entanglement phenomenon as 
described by conventional quantum mechanics.
The search for correlations between entangled particles started in the 1970's by several groups \cite{FS}.. Typically, the 
particles involved were photon pairs that had been emitted in electronic transitions of calcium or mercury atoms and travelled 
in opposite directions with entangled polarizations. Each photon impacted on a polarization analyzer and the investigators
could verify in this way that the correlations predicted by the quantum entanglement were confirmed massively, violating
the Bell's inequalities in some settings \cite{Bell}. However, the investigators were not completely satisfied as the 
orientations of the analyzers were fixed before the photons were emitted, and they were afraid that somehow relevant 
information could be exchanged between the analyzers\footnote{In our opinion, the possibility that the two analyzers could 
talk to each other is by far more astonishing and revolutionary than the quantum correlations themselves.}.
 
Finally, in 1981 the French group of Aspect and collaborators succeeded in switching the analyzers orientations faster than any 
light signal could be interchanged between them \cite{Aspect}. The results were the same as with the less sophisticated equipments
but this time most researchers felt comfortable and the violation of the Bell's inequalities was finally established, ruling 
out in the process realistic local hidden variable interpretations of quantum mechanics 
forever\footnote{Not everybody was satisfied, however, as there were complaints because the switchings of the analyzers 
orientations were periodic rather than random. Finally, in 2015 the Dutch-Spanish group of Hensen and collaborators 
succeded in obtaining a so-called loophole-free violation of a Bell inequality using entangled electron spins \cite{Hensen}.}. 
In 2003 Leggett \cite{Leggett} proposed an inequality for testing a class
of realistic interpretations of quantum mechanics involving nonlocal hidden variables; that is, maintaining well
defined individual properties of the particles. Experimental violations of the Leggett inequality were found in 2007
\cite{GPB} using polarization states of photons, thereby discarding this class of realistic interpretations. Moreover, 
in 2010 the Leggett inequality was violated again, using the orbital angular momentum states of photons \cite{Romero}.

As stated in section 2, Scan Quantum Mechanics predicts, in agreement with conventional 
quantum mechanics, that the exact values of the polarizations of the entangled
particles are random but strictly correlated with each other, no matter the distance that separates the two particles.
However, in our proposal each photon switches very rapidly and randomly among the available 
polarization states, the switchings of the two photons being exactly synchronized with each other. 
In these experiments both switchings stop when the polarization of one of the photons is measured, 
whereas in conventional quantum mechanics each photon only finds itself
in a well defined polarization state when it is measured. That is, in Scan Quantum Mechanics when the polarization
of one of the photons is measured, then the other photon instantly acquires the correlated polarization without the need
of any measurement, whereas in conventional quantum mechanics the other photon does not acquire any polarization until it
is measured, no matters the time passed by, even tough the result of the measurement is known beforehand because it must 
be correlated with the result of the first photon. This behaviour we find extremely weird.

\subsection{Possible experiments to probe Scan Quantum Mechanics}

\vskip .2in

In Scan Quantum Mechanics finding the dividing line between quantum and classical behaviour amounts to finding the 
values of the critical quantum inertia $I_{cr}$ for all the observables of a system, including its mass. 
Adding constituents to a quantum system is only a particular case, although perhaps the most effective one, for 
increasing its quantum inertia $I_q$, via the masses and interactions of the new constituents. 
But the mass or number of constituents are not the only source of quantum inertia of a physical system, as we explained
in section 2, and all the interactions coming from the environment must also contribute: temperature, gravitational 
strength, electric and magnetic fields and collisions with all kinds of matter and radiation. In addition, 
different interactions affect the observables in different ways and different observables are
expected to have different values for the critical inertia $I_{cr}$. 

For fixed environmental conditions, adding constituents to a quantum system until $I_q$ reaches or surpasses the 
corresponding critical value, $I_q \geq I_{cr}$, results in the cessation of the system's quantum jumps between the two
(or more) trajectories in interferometry experiments: the system has become classical and follows a well defined
trajectory. As a consequence, no interference pattern can be created by repeating the experiment with a large number of 
identical systems, unless one or more of the external conditions change, for example a decrease in temperature. 
We conclude, therefore, that it should be technically possible to determine (or estimate at least) the critical 
quantum inertia $I_{cr}$ associated to the mass of physical systems for given external conditions, 
by performing interferometry experiments probing large molecules, in the lines reported in ref. \cite{Hornberger}, 
or nanocrystals to which the experimentalists could add or subtract atoms one by one. 

Once the dividing line had been found between quantum and classical behaviour regarding the mass of the physical 
systems, for fixed environmental conditions, the next interferometry experiments must be performed using borderline 
systems (i.e. very close to the dividing line), varying slightly one of those conditions. For example,  
increasing and decreasing the temperature to test its effect on the quantum inertia $I_q$, as well as 
testing the reversibility of the process, with the possibility of finding a critical temperature $T_{cr}$. A technical
difficulty will always appear, as we already mentioned in section 2. Namely, the hardness to distinguish between 
the influence of temperature on the quantum inertia and the thermal
fluctuations, which also spoil quantum coherence. In this respect, it is already known \cite{Hackermuller} that at 1500 K 
the fullerens still behave as quantum waves while they are indistinguishable from classical particles when they are close 
to 3000 K. It is also known that the critical temperature for effective quantum-to-classical transition decreases with 
increasing size \cite{Joos} \cite{Hornberger06}, and it is conjectured that thermal decoherence should be avoidable 
for particles with masses up to $10^9$ amu by cooling them to their vibrational ground state, below 77 K (see \cite{Hornberger} 
and references therein).

Unfortunately one cannot do similar tests increasing and decreasing the gravitational strength substantially, 
although experiments could be performed (perhaps) in increasingly accelerated devices. Nevertheless, 
with regard to gravitational strength, one can easily deduce that stars lighter than neutron stars, like white dwarfs, 
are uncapable to stop the switching among positions of the electrons in atoms, since otherwise these astrophysical objects 
would turn very quickly into neutron stars, emitting enormous amounts of $\gamma$-ray synchrotron radiation in the process. 

The effect of magnetic and electric fields on $I_q$ should also be tested in interferometry experiments using borderline 
systems. In addition, one can consider other tests, for example, the Stern-Gerlach experiment (1922) in which silver atoms 
passing through non-uniform magnetic fields split in two trajectories independently of the orientation of the magnetic field. 
This type of experiments should be repeated in order to search for two effects. First, lowering as much as possible the strength 
of the magnetic field one should try to detect any thresholds pointing towards the existence of a critical $I_{cr}$. Second, 
one should try to determine how long it takes for the atom to come back to the initial 
superposition, after it passed through the magnetic field, if let alone unperturbed (not if the atom passes through another 
magnetic field pointing in another direction, for this we already know the answer). For if one could test that the atoms come 
back to the initial superposition, explained by Scan Quantum Mechanics simply as a release of quantum inertia, so that
$I_q \leq I_{cr}$, then the corresponding $I_{cr}$ could be estimated, and the reversibility of the process would be demonstrated.
However, in practice, long time spans and long distances might be involved for the release of the necessary amount of $I_q$ in 
order to accomplish the reversibility of the superposition. In this event the reversibility would be impossible to test in 
laboratory settings.


Another experiment is to estimate the time it takes for a quantum system to liberate 
a subsystem from an entanglement bound, that is to dissolve the entanglement, after the partner  
has been detected, simply by releasing enough quantum inertia to the environment. Observe that this has nothing to do with disentangling
a system before any one of the particles has been detected, as it seems to happen in entanglement sudden death due mainly to decoherence
(with perhaps some contribution from the quantum inertia). In the usual experimental settings, where the entangled particles or atoms
travel very fast in opposite directions, we pointed out in section 3 that perhaps very large, even interstellar, 
distances might be necessary to test 
this effect. In other words, like in the case of the Stern-Gerlach experiments, it could be impossible in practice to test that once one 
of the particles has been detected, the entanglement with the other particle finally disappears due to the release of quantum inertia. 
However, only by performing these experiments we would be able to put bounds on the corresponding values of $I_{cr}$ and demonstrate the 
reversibility of such `measurements', in full disagreement with conventional quantum mechanics.

In refs. \cite{Hornberger}, \cite{Bassi}, \cite{Bahrami} a number of experimental settings and proposals are described
in order to probe the quantum-to-clasical transition. Although these experiments are designed to test decoherence theory or
collapse models, they can be applied as well to test Scan Quantum Mechanics, i.e. to test the existence of quantum inertia with 
critical values $I_{cr}$, and eventually the associated critical temperatures $T_{cr}$. For example, as was mentioned in section 2,
collisional decoherence has been studied quantitatively using a Talbot-Lau interferometer \cite{Horn2003} by gradual
admision of different gases into the vacuum chamber and it was found that at room temperature a single scattering event
per molecule of fullerens suffices to fully destroy the interference.

\vskip .4in

\section{Conclusions and Final Remarks}

\vskip .2in

In this article we have presented the essential features of Scan Quantum Mechanics, a novel interpretation with respect to the 
meaning of the wave function and the superposition of states. We postulate a crucial property of all physical systems called 
quantum inertia $I_q$, that increases whenever a constituent is added to the system and also when the system is 
perturbed with all kinds of interactions and energies. 

In our proposal the quantum systems are in well defined states at each instant, but switch randomly and very fast among 
the available states described by the wave function, producing an effective apparent superposition of states. 
We quantify the time spent in each state in terms of a unit called switching time $t_S$, which must depend on 
the system's quantum inertia $I_q$. We have proposed a specific expression of $t_S$ as function of $I_q$ in Eq.(1), and
we have also proposed a lower bound relating the indeterminacy on the quantum inertia $\Delta I_q$ with the corresponding 
indeterminacy on the switching time $\Delta t_S$.

The switching among the available states is a natural fact of the quantum world, without the need of any specific cause, 
and only stops when the quantum inertia $I_q$ reaches or surpasses a critical value $I_{cr}$, 
that may be different for each observable. In this case the switching time becomes infinite, $t_S \rightarrow \infty $, 
signalizing the end of the superposition, and the system stabilizes in one specific state with well defined physical 
properties of the observable at hand. Thus for Scan Quantum Mechanics the origin of classicality, the dividing line between 
quantum and classical behaviour, is simply described by the relation $I_q \geq I_{cr}$. As a result, decreasing the load of 
quantum inertia of a system below $I_{cr}$ for a given observable (whenever this is possible) should revert
the process. This reversibility does not imply time reversal symmetry, however, as the switchings of the systems
among the available states are random. If there is more than one value of $I_{cr}$, corresponding to different observables, 
then quantum and classical behaviour should co-exist for some hybrid systems, especially at very low temperatures and weak 
gravitational fields, as it actually happens in mesoscopic systems.  

Positions in space and momenta are treated in a similar way: the systems switch among the available positions, 
not necessarily all of them, but a large subset at least. As 
a result, for quantum systems continuous classical trajectories do not exist, but quantum jumps are the rule. 
A very interesting case is when $I_q \geq I_{cr}$ where $I_{cr}$ is the critical inertia for the quantum jumps. 
Then the system is unable to perform them and moves like a classical object. If this happens to electrons 
in atoms due to, for example, the action of very strong gravitational fields, then the atoms will look like tiny 
planetary systems, with the electrons emitting $\gamma$-ray synchrotron radiation while falling into the nuclei.
This non-thermal emission from astrophysical objects with strong gravitational fields, like neutron stars or black holes,
should also be taken into account as an efficient mechanism for the conversion of atoms into neutrons in neutron stars 
and in accretion discs of some black holes, producing sudden explosions of gamma rays in all directions (less intense, 
however, than the `canonical' gamma ray bursts). Obviously, the gravitational strength of white dwarfs is not enough 
to produce this mechanism, as these stars would turn into neutron stars very quickly otherwise. Therefore, the gravitational
strength of white dwarfs provide a lower bound for the critical gravitational strength necessary to ignite this mechanism.
That is, $G_{cr} > G_{wd}$, where $wd$ stands for white dwarfs.

As to quantum entanglement, it requires exact synchronization between the switchings of the systems involved. As a result
a quantum system cannot get entangled with a classical one unless the quantum inertia of the latter is very close to the 
critical value and, somehow, it can be lowered even more through the contact with the quantum system. Once one partner 
stops switching, for example due to a measurement, the switching stops instantly for all the entangled partners. The reason 
is that the quantum system acts as a whole, independently of the physical locations of its parts in spacetime. This produces 
the freezing of the entangled property for all the partners involved, but not forever, for as soon as they could release
enough quantum inertia, the partners would resume the switching activity, liberating themselves from the 
entanglement bound.
 
We have discussed Scan Quantum Mechanics versus several other interpretations of quantum mechanics. Most of them treat 
classical macroscopic objects as quantum objects, with the corresponding wave functions that describe their behaviour. 
This is due to the fact that these interpretations lack a clear mechanism to draw the dividing line between these
two kinds of objects, between the quantum and the classical worlds. 

As regards the domain of applicability of the quantum-mechanical equations, in Scan Quantum Mechanics trying to apply 
Schr\"odinger's equation to a classical system is as misleading as trying to apply the ballistic parabolic equations of 
motion to a bullet after it has hit the ground. Classical systems have hit the ground, i.e. have surpassed the critical
quantum inertia, and as a consequence their wave functions (if one bothers to compute them) do not describe their physical
reality nor does Schr\"odinger's equation make any sense for them. 

One may wonder what causes the quantum switching, in the first place. As we said, 
these are facts of the quantum world and nothing special causes them, not a quantum potential or any forces, 
but these features are simply a primary, basic behaviour of the quantum systems, like randomness. Similarly, 
one might ask what causes a massless particle to move at the maximum speed c in vacuum. Certainly there are no hidden 
potentials in the vacuum pushing massless particles on the back in order to reach the maximum speed. This is simply
a basic property of nature. More meaningful, in these cases, is turning the question upside down: 
what causes massive particles to move slower than c ? This we can answer in a more intuitive way by invoking the 
inertial effects caused by the masses. In the same way, we should turn upside down the question about the 
quantum switching. What causes the classical systems not to switch between different states, i.e. not to have 
superpositions, and present well defined properties as a result? The answer given by Scan Quantum Mechanics is quantum 
inertia $I_q$ reaching or surpassing a critical value $I_{cr}$. 

We assume that the wave function evolves unitarily with the Schr\"odinger equation, like in conventional quantum mechanics.
But our interpretation is more asymmetric with respect to space and time since we postulate that quantum systems 
are switching states as time passes by. The r\^ole of time is therefore more central in our proposal than in conventional 
quantum mechanics. The reason could be that there are two levels of physical reality: the more fundamental, deeper quantum 
level, with time but no space, and the spacetime level at the surface, the former imposing its rules on the latter,
producing nonlocal effects such as quantum jumps and entanglement. For this reason too we think that the effect of quantum
inertia in the physical systems remains deep at the quantum level, not showing up in the spacetime dynamics in the form of 
a non-linear term in the hamiltonian of the Schr\"odinger equation.

We have reviewed the most relevant past and present experiments in quantum mechanics, in the light of the 
more intuitive description offered by Scan Quantum Mechanics. We also have proposed some experiments to be performed 
in order to determine the critical value $I_{cr}$ for some observables, perhaps even an associated critical temperature 
$T_{cr}$, and the reversibility of the dividing line between the quantum and the classical worlds. These are experimental
input in this interpretation, like the masses of elementary particles in the Standard Model of particle physics. 
As a matter of fact,
many experiments that have been performed, or proposed, in the last years by a number of groups in order to test the
quantum-to-clasical transition \cite{Hornberger}, \cite{Horn2003}, \cite{Bassi}, \cite{Bahrami}, apply similarly to
determine (or estimate) the critical quantum inertia $I_{cr}$ and the critical temperature $T_{cr}$.
Furthermore, we think that mesoscopic systems in general as well as
physical phenomena at very low temperatures, such as superconductivity, superfluidity 
and Bose-Einstein condensates must be reanalyzed in the light of this interpretation, i.e. taking into account the
possibility of existence of quantum inertia with a critical value $I_{cr}$, and the associated critical $T_{cr}$.

Finally, let us notice that, despite the asymmetry with respect to space and time, the essential ideas of Scan Quantum 
Mechanics can be transferred straightforwardly to Quantum Field Theory, in which case the probabilities are implemented 
by switchings and quantum jumps of the particles, the quantum inertia giving rise to natural ultraviolet cutoffs.
This brings about a very intriguing possibility: does quantum gravity really exist? For if strong enough gravitation 
disables quantum mechanics via quantum inertia, then there is no much room left for quantum gravity.



\vskip .5in
\noindent
{\bf \large Acknowledgements:}
\vskip .2in
\noindent
I am grateful to Alberto Galindo, Jaime Julve, Salvador Miret and Pablo Villareal for suggestions 
in order to improve the presentation of this article. I also thank Mar\'\i a Victoria Fonseca,
Jose Gonz\'alez Carmona, Felipe Llanes, Salvador Robles P\'erez, Pilar Ruiz Lapuente and Alfredo Tiemblo 
for useful conversations. This research was supported partially by the project FIS2012-38816,
from the Spanish Ministerio de Econom\'\i a y Competitividad.

\bibliography{REFS}

\end{document}